\documentclass[]{article}
\pdfoutput=1
\usepackage{lmodern}
\usepackage{amssymb,amsmath}
\usepackage{ifxetex,ifluatex}
\usepackage{fixltx2e} % provides \textsubscript
\ifnum 0\ifxetex 1\fi\ifluatex 1\fi=0 % if pdftex
  \usepackage[T1]{fontenc}
  \usepackage[utf8]{inputenc}
\else % if luatex or xelatex
  \ifxetex
    \usepackage{mathspec}
  \else
    \usepackage{fontspec}
  \fi
  \defaultfontfeatures{Ligatures=TeX,Scale=MatchLowercase}
\fi
\IfFileExists{upquote.sty}{\usepackage{upquote}}{}
\IfFileExists{microtype.sty}{%
\usepackage{microtype}
\UseMicrotypeSet[protrusion]{basicmath} % disable protrusion for tt fonts
}{}
\usepackage[margin=1in]{geometry}
\usepackage{hyperref}
\hypersetup{unicode=true,
            pdftitle={Correct power for cluster-randomized difference-in-difference trials with loss to follow-up},
            pdfauthor={Jonathan Moyer and Ken Kleinman},
            pdfborder={0 0 0},
            breaklinks=true}
\usepackage{graphicx,grffile}
\makeatletter
\def\maxwidth{\ifdim\Gin@nat@width>\linewidth\linewidth\else\Gin@nat@width\fi}
\def\maxheight{\ifdim\Gin@nat@height>\textheight\textheight\else\Gin@nat@height\fi}
\makeatother
\setkeys{Gin}{width=\maxwidth,height=\maxheight,keepaspectratio}
\IfFileExists{parskip.sty}{%
\usepackage{parskip}
}{% else
\setlength{\parindent}{0pt}
\setlength{\parskip}{6pt plus 2pt minus 1pt}
}
\ifx\paragraph\undefined\else
\let\oldparagraph\paragraph
\renewcommand{\paragraph}[1]{\oldparagraph{#1}\mbox{}}
\fi
\ifx\subparagraph\undefined\else
\let\oldsubparagraph\subparagraph
\renewcommand{\subparagraph}[1]{\oldsubparagraph{#1}\mbox{}}
\fi

\let\rmarkdownfootnote\footnote%
\def\footnote{\protect\rmarkdownfootnote}

\usepackage{titling}

  \title{Correct power for cluster-randomized difference-in-difference trials
with loss to follow-up}
    \pretitle{\vspace{\droptitle}\centering\huge}
  \posttitle{\par}
    \author{Jonathan Moyer and Ken Kleinman}
    \preauthor{\centering\large\emph}
  \postauthor{\par}
    \date{}
    \predate{}\postdate{}
  
\usepackage{booktabs}
\usepackage{longtable}
\usepackage{array}
\usepackage{multirow}
\usepackage{wrapfig}
\usepackage{float}
\usepackage{colortbl}
\usepackage{pdflscape}
\usepackage{tabu}
\usepackage{threeparttable}
\usepackage{threeparttablex}
\usepackage[normalem]{ulem}
\usepackage{makecell}
\usepackage{xcolor}

\usepackage{amsmath}
\usepackage{booktabs}
\usepackage{bm}
\usepackage{amssymb}
\usepackage{tikz}
\usetikzlibrary{matrix,fit,decorations.pathreplacing,calc,graphs,decorations.markings,intersections,positioning}
\usepackage{changepage}

\begin{document}
\maketitle
\begin{abstract}
Cluster randomized trials with measurements at baseline can improve
power over post-test only designs by using difference in difference
designs. However, subjects may be lost to follow-up between the baseline
and follow-up periods. While equations for sample size and variance have
been developed assuming no loss to follow-up (``cohort'') and completely
different subjects at baseline and follow-up (``cross-sectional'')
difference in difference designs, equations have yet to be developed
when some subjects are observed in both periods (``mixture'' designs).
We present a general equation for calculating the variance in difference
in difference designs and derive special cases assuming loss to
follow-up with replacement of lost subjects and assuming loss to
follow-up with no replacement but retaining the baseline measurements of
all subjects. Relative efficiency plots, plots of variance against
subject autocorrelation, and plots of variance by follow-up rate and
subject autocorrelation are used to compare cohort, cross-sectional, and
mixture approaches. Results indicate that when loss to follow-up to
uncommon, mixture designs are almost as efficient as cohort designs with
a given initial sample size. When loss to follow-up is common, mixture
designs with full replacement maintain efficiency relative to cohort
designs. Finally, our results provide guidance on whether to replace
lost subjects during trial design and analysis.
\end{abstract}

\hypertarget{introduction}{%
\section{1. Introduction}\label{introduction}}

Cluster randomized trials (CRTs) are trials in which groups or clusters
of individuals are randomly assigned to treatment conditions. This is in
contrast to an individually randomized trial (IRT), in which individuals
themselves are randomly assigned to treatment conditions. A trial may be
be conducted as a CRT for reasons of administrative convenience or to
avoid cross contamination (Hayes and Moulton 2017). However, the
tendency of measurements within a given cluster to be correlated that
must be accounted for when when designing and analyzing a CRT. Failing
to account for this will typically increase the Type 1 error rate.

The ``parallel arm'' CRT design consists of measuring the response after
implementation of the treatment. However, it is common to record
measurements of the response at baseline as well as after treatment
implementation. In this situation, analysis can focus on the difference
in differences (DID) - that is, the difference in change between
treatment and control groups. Examples include studies in improving
stroke rehabilitation (Strasser et al. 2008), dating violence prevention
(Miller et al. 2012), and community-based modification of harmful gender
norms (Pettifor et al. 2018). Use of the DID design can result in
improvements to power (Rutterford, Copas, and Eldridge 2015).

Two classes of DID design have long been recognized (Murray and Hannan
1990). The first is the cohort design, in which the same subjects are
measured at baseline and at follow-up. The second is the cross sectional
design, in which subjects within a cluster at follow-up are different
from those at baseline. A single expression has been developed for the
variance of the DID estimator that incorporates aspects of both cohort
and cross-sectional designs (Feldman and McKinlay 1994).

Mixtures of the two designs are also possible. In this scenario,
individuals measured at baseline and lost to follow-up may be replaced.
Thus complete observations are available for some individuals, those
lost to follow-up have only baseline measurements, and the replacements
have only follow-up measurements. An example is the REDUCE MRSA trial,
where participating hospitals were randomly assigned to one of three
strategies to prevent health-care associated infections (Huang et al.
2013). In this trial, virtually all subjects lost to follow-up were
replaced with new subjects, but some subjects were present at baseline
and follow-up. Previous work (Teerenstra et al. 2012), (Feldman and
McKinlay 1994) mentions that the formulation for the DID estimator
variance and design effect can be applied to mixtures of cohort and
cross-sectional designs but does not provide further details.

Mixtures of the two designs could be considered from the perspective of
missing data, which is a common occurence in CRTs (Fiero et al. 2016).
That is, individuals present only at baseline have missing values at
follow-up, while individuals present only at follow-up have missing
values at baseline. The cross-sectional design could be considered an
extreme case in which all individuals at baseline and follow-up are
respectively missing their follow-up and baseline values. Another
extreme case of a ``mixture'' design is when individuals are lost to
follow-up and not replaced. In an IRT, a DID analysis must necessarily
omit those lost to follow-up. However, in a CRT, those lost to follow-up
still contribute to the cluster means at baseline.

The goal of this article is to develop an approach to calculating the
DID estimator variance and design effect for the mixture design,
including replacement of those lost to follow-up or no replacement at
all. In section 2 we review the model used for the DID design. In
section 3 we present variations to the variance of the DID estimator
assuming replacement of drop-outs and consider power and sample size
calcualtions. In section 4, we compare the variance of DID estimators
using relative efficiency plots, plots of predicted variance against
subject autocorrelation \(\rho_S\), and plots indicating which DID
design predicts the smallest variance given pairs of follow-up rates.
Finally, in section 5 we offer further discussion and concluding
thoughts.

\hypertarget{did-model-review}{%
\section{2. DID Model Review}\label{did-model-review}}

Let \(y_{itjk}\) be a continuous outcome for individual \(k\)
\((k = 1, \dots, K_{itj})\) in cluster \(j\) \((j = 1, \dots, J)\) and
treatment group \(i\) \((i = 1,2)\) at time \(t\) \((t = 1,2)\), where
\(J\) is the total number of clusters per arm and \(K_{itj}\) is the
number of individuals in cluster \(ij\) at time \(t\). Let \(A_i\) and
\(T_t\) be indicators for arm (\(A_1 = 0\) for control, \(A_2 = 1\) for
treatment) and time (\(T_1 = 0\) for baseline, \(T_2 = 1\) for
follow-up). Then the model for \(y_{itjk}\) is
\begin{equation} \label{eq:yitjk}
y_{itjk} = \beta_0 + \beta_1 A_i + \beta_2 T_t + \beta_3 A_i T_t + C_{itj} + (CT)_{itj} + S_{itjk} + (ST)_{itjk}
\end{equation}

Fixed effects include the first four terms of equation \ref{eq:yitjk},
where \(\beta_0\) represents the mean of the control group at baseline,
\(\beta_1\) is the mean baseline difference between treatment and
control groups, \(\beta_2\) represents the mean difference between
follow-up and baseline values for the control group, and \(\beta_3\) is
mean differential change between treatment and control. The parameter
\(\beta_3\) corresponds to the DID.

Random effects are represented by the last four terms of equation
\ref{eq:yitjk}. They are assumed to be mutually independent and normally
distributed as follows:

\begin{center}
\begin{tabular}{ll}
\toprule
Term & Distribution \\
\midrule
$C_{itj}$ & $N(0,\sigma_{C}^2)$ \\
$(CT)_{itj}$ & $N(0,\sigma_{CT}^2)$ \\
$S_{itjk}$ & $N(0,\sigma_{S}^2)$ \\
$(ST)_{itjk}$ & $N(0,\sigma_{ST}^2)$ \\
\bottomrule
\end{tabular}
\end{center}

The terms \(C_{itj}\) and \((CT)_{itj}\) represent time-invariant and
time-varying effects due to cluster, respectively. Each cluster will
have one value for \(C_{itj}\) and two values for \((CT)_{itj}\)
corresponding to baseline and follow-up. The terms \(S_{itjk}\) and
\((ST)_{itjk}\) represent time-invariant and time-varying effects due to
subject, respectively. As with clusters, each subject will have one
value for \(S_{itjk}\) and two values for \((ST)_{itjk}\) corresponding
to baseline and follow-up. Indeed, in the common case where each subject
is measured once per time point, the term \((ST)_{itjk}\) serves as
residual error. The \((ST)_{itjk}\) notation is used here to be
consistent with prior work by others.

The intracluster correlation coefficient (ICC), \(\rho\), is a measure
of the correlation between measurements within a cluster. Often used in
power analysis for CRTs, in the DID setting \(\rho\) is defined as
follows: \begin{equation}
\rho = \frac{\sigma_C^2 + \sigma_{CT}^2}{\sigma_C^2 + \sigma_{CT}^2 + \sigma_S^2 + \sigma_{ST}^2}
\end{equation}

Total cluster and subject variances are partitioned into time-invariant
and time-varying components through cluster and subject autocorelation
coefficients \(\rho_C\) and \(\rho_S\), respectively. The parameter
\(\rho_C\) indicates the correlation between baseline and follow-up
means for a cluster, while \(\rho_S\) indicates the correlation between
baseline and follow-up values for an individual, conditional on cluster
random effects. In terms of the random effect variances, the
autocorrelation parameters are defined as:
\begin{align} \label{eq:autocorr}
\rho_C &= \frac{\sigma_{C}^2}{\sigma_{C}^2 + \sigma_{CT}^2} \nonumber \\
\rho_S &= \frac{\sigma_{S}^2}{\sigma_{S}^2 + \sigma_{ST}^2}
\end{align}

The type of DID design - cross-sectional, cohort, or mixture - is
reflected in the \(\rho_S\) parameter. If \(\rho_S = 0\), then there is
no correlation between baseline and follow-up subject measurements and
the design is effectively cross-sectional. Values of \(\rho_S > 0\) only
have meaning as a cohort design. Assuming some fraction of subjects are
measured twice with autocorrelation \(\rho_S\) and the rest measured
only once, the subject autocorrelation for a mixture design, defined as
\(\rho_S^*\), will also be greater than 0 but less than \(\rho_S\).
(Teerenstra et al. 2012) One aim of section 3 below is to characterize
to what degree \(\rho_S^*\) is less than \(\rho_S\).

Let \(\bar{y}_{it}\) be the average of the realized observations of
group \(i\) at time \(t\). Assuming equal number of clusters per arm and
an equal number of subjects across clusters (i.e, \(K_{itj} = K\)), the
estimator of the interaction term \(\beta_3\) in equation \ref{eq:yitjk}
- denoted as \(\hat{\beta}_3\) - can be represented in terms of
\(y_{itjk}\) as follows: \begin{align}
  \hat{\beta}_3 &= (\bar{y}_{22} - \bar{y}_{21}) - (\bar{y}_{12} - \bar{y}_{11}) \nonumber \\
  &= \frac{1}{JK}\Bigg[\sum_{j=1}^J\sum_{k=1}^K y_{11jk} - \sum_{j=1}^J\sum_{k=1}^K y_{12jk} - \sum_{j=1}^J\sum_{k=1}^K y_{21jk} + \sum_{j=1}^J\sum_{k=1}^K y_{22jk} \Bigg]
\end{align}

The variance of this estimator is

\begin{align} \label{eq:vardid}
Var(\hat{\beta}_3) &= 4\Big[\frac{\sigma_{CT}^2}{J} + \frac{\sigma_{ST}^2}{JK}\Big] \nonumber \\
                   &= 4\Big[\frac{\sigma_{CT}^2}{J} + \frac{(1 - \rho_S)(\sigma_{S}^2+\sigma_{ST}^2)}{JK}\Big]
\end{align}

As discussed above, for the cross-sectional design \(\rho_S = 0\). In
this case, equation \ref{eq:vardid} becomes

\begin{equation} \label{eq:vardidcross}
Var(\hat{\beta}_3) = 4\Big[\frac{\sigma_{CT}^2}{J} + \frac{\sigma_{S}^2+\sigma_{ST}^2}{JK}\Big]
\end{equation}

\hypertarget{ltf-variations-on-did}{%
\section{3. LTF Variations on DID}\label{ltf-variations-on-did}}

We present variations on the DID design incorporating loss to follow-up
(LTF) and potentially replacement. In all cases, it is assumed that data
are missing completely at random (MCAR) as defined by Rubin (Rubin
1976).

\hypertarget{general-case}{%
\subsection{3.1 General case}\label{general-case}}

Let \(L_i\) and \(G_i\) be the numbers lost to and gained at follow-up
per cluster in arm \(i\). It is possible for \(G_i\) to be greater than
\(L_i\). Assume the initial number of subjects in each cluster is \(K\).
With respect to \(K\), the loss to follow-up and gain at follow-up rates
per cluster in arm \(i\) \(\lambda_i\) and \(\gamma_i\), respectively,
are given by \(\lambda_i = L_i/K\) and \(\gamma_i = G_i/K\).
Furthermore, let
\(\eta = \frac{1-\lambda_1}{1-\lambda_1+\gamma_1}+\frac{1-\lambda_2}{1-\lambda_2+\gamma_2}-1\).
Then the subject autocorrelation modified for losses to and gains at
follow-up \(\rho_S^*\) is given by (see the appendix)
\begin{equation} \label{rhoSgeneral}
\rho_S^* = \rho_S - \frac{1}{4}\Big[\frac{1}{1-\lambda_1+\gamma_1} + \frac{1}{1-\lambda_2+\gamma_2} - 2\frac{\eta\sigma_S^2 + \sigma_{ST}^2}{\sigma_S^2+\sigma_{ST}^2}\Big]
\end{equation}

This allows the variance of the DID estimator modified by gains and
losses to follow-up to be represented as follows:
\begin{align} \label{vardidgeneral}
Var(\hat{\beta}_3) &= 4\Bigg[\frac{\sigma_{CT}^2}{J} + \frac{(1-\rho_S^*)(\sigma_S^2+\sigma_{ST}^2)}{JK}\Bigg]
\end{align}

Setting \(\lambda_i\) and \(\gamma_i\) to various quantities results in
variations on the variance of the DID estimator, as discussed below.

\hypertarget{ltf-with-replacement-and-without-replacement-variations}{%
\subsection{3.2 LTF with replacement and without replacement
variations}\label{ltf-with-replacement-and-without-replacement-variations}}

The mixture design discussed in section 2 consists of complete loss to
follow-up with full-replacement. In this case, in equation
\ref{rhoSgeneral} \(\lambda_i = \gamma_i\) and
\(\eta = 1 - \lambda_1 - \lambda_2\). Then
\(\rho_S^* = (1 - \bar\lambda)\rho_S\), where
\(\bar\lambda = \frac{\lambda_1+\lambda_2}{2}\) is the mean loss to
follow-up rate across both arms. The variance of the DID estimator with
complete loss to follow-up with full replacement, \(\beta_{3,rep}\), is
therefore \begin{align} \label{vardidwrep}
Var(\hat{\beta}_{3,rep}) &= 4\Bigg[\frac{\sigma_{CT}^2}{J} + \frac{(1-\rho_S^*)(\sigma_S^2+\sigma_{ST}^2)}{JK}\Bigg] \nonumber \\
                   &= 4\Bigg[\frac{\sigma_{CT}^2}{J} + \frac{(1-(1 - \bar\lambda)\rho_S)(\sigma_S^2+\sigma_{ST}^2)}{JK}\Bigg]
\end{align}

Setting \(\gamma_i = 0\) in equation \ref{rhoSgeneral} and assuming
\(0 < \lambda_i < 1\) results in a loss to follow-up with no replacement
scenario. In this case, \(\rho_S^*\) is given by
\(\rho_S^* = \rho_S - \frac{1}{4}\Big[\frac{\lambda_1}{1-\lambda_1}+\frac{\lambda_2}{1-\lambda_2}\Big]\).
The variance of the DID estimator with complete loss to follow-up with
full replacement, \(\beta_{3,ltf}\), is given by
\begin{align} \label{vardidltfnorep}
Var(\hat{\beta}_{3,ltf}) &= 4\Bigg[\frac{\sigma_{CT}^2}{J} + \frac{(1-\rho_S^*)(\sigma_S^2+\sigma_{ST}^2)}{JK}\Bigg] \nonumber \\
                   &= 4\Bigg[\frac{\sigma_{CT}^2}{J} + \frac{\Big(1-\Big\{\rho_S - \frac{1}{4}\Big[\frac{\lambda_1}{1-\lambda_1}+\frac{\lambda_2}{1-\lambda_2}\Big]\Big\}\Big)(\sigma_S^2+\sigma_{ST}^2)}{JK}\Bigg]
\end{align}

\hypertarget{power-calculation}{%
\subsection{3.3 Power Calculation}\label{power-calculation}}

The power to detect DID effect \(\beta_3\) with \(J\) clusters per arm
is given by \begin{equation}
Power = \Phi_{t,2(J-1)}\Bigg(\frac{\beta_3}{\sqrt{Var(\hat{\beta}_3)}} - t_{\alpha/2,2(J-1)}\Bigg)
\end{equation}

where \(\Phi_{t,\nu}\) is the cumulative \(t\)-distribution with \(\nu\)
degrees of freedom.

Using one of the subject autocorrelations described previously, one can
solve for the minimum number of clusters per arm needed to attain a
given power at a given level of significance \(\alpha\).
\begin{equation*}
J = 4\Bigg(\sigma_{CT}^2+\frac{(1-\rho_S)(\sigma_S^2+\sigma_{ST}^2)}{K}\Bigg)\Bigg(\frac{t_{\beta,2(J-1),\phi} + t_{\alpha/2,2(J-1)}}{\beta_3}\Bigg)^2
\end{equation*}

where non-centrality parameter
\(\phi = \hat\beta_3/\sqrt{Var(\hat\beta_3)}\). In practice, power
calculation software, such as that provided by the \texttt{clusterPower}
package in R, can also compute power and other quantities of interest
(Kleinman, Moyer, and Reich 2017).

\hypertarget{example-sample-size-determination}{%
\subsection{3.4 Example Sample Size
Determination}\label{example-sample-size-determination}}

To illustrate the use of the preceding equations, we proceed as if we
were planning an intervention to determine the effects of the dating
violence preventon program described by Miller et al (2012). One goal of
this study was to improve high school aged male athletes' inclination to
intervene when witnessing abusive behavior, as measured on an 8-item
scale. The study was a cohort design with 16 high schools as the
clustering units. At baseline, approximately 125 students per school
participated. The analysis was performed on complete case data, but the
authors report follow-up rates of 84\% and 95\% in the treatment and
control groups, respectively. The adjusted mean difference in intention
to intervene change was observed to be 0.12, meaning intervention youth
increased intent to intervene by 0.12 points on average relative to
control youth (\(\hat\beta_3 = 0.12\)). Correspondence with the authors
indicated that the values of \(\sigma_C^2\), \(\sigma_{CT}^2\),
\(\sigma_S^2\), and \(\sigma_{ST}^2\) were 0.0218, 0.0047, 0.3342, and
0.2567, respectively. This produces values of 0.0429, 0.8226, and 0.5656
for \(\rho\), \(\rho_C\), and \(\rho_S\), respectively.

In our intervention, suppose we have 30 schools participating (15 per
arm) and want to know the minimum number of subjects per school needed
to attain 80\% power at a 5\% level of significance. Using the utilities
in the R package \texttt{clusterPower}, the power for each design can be
computed over a number of subjects.

\begin{center}\includegraphics{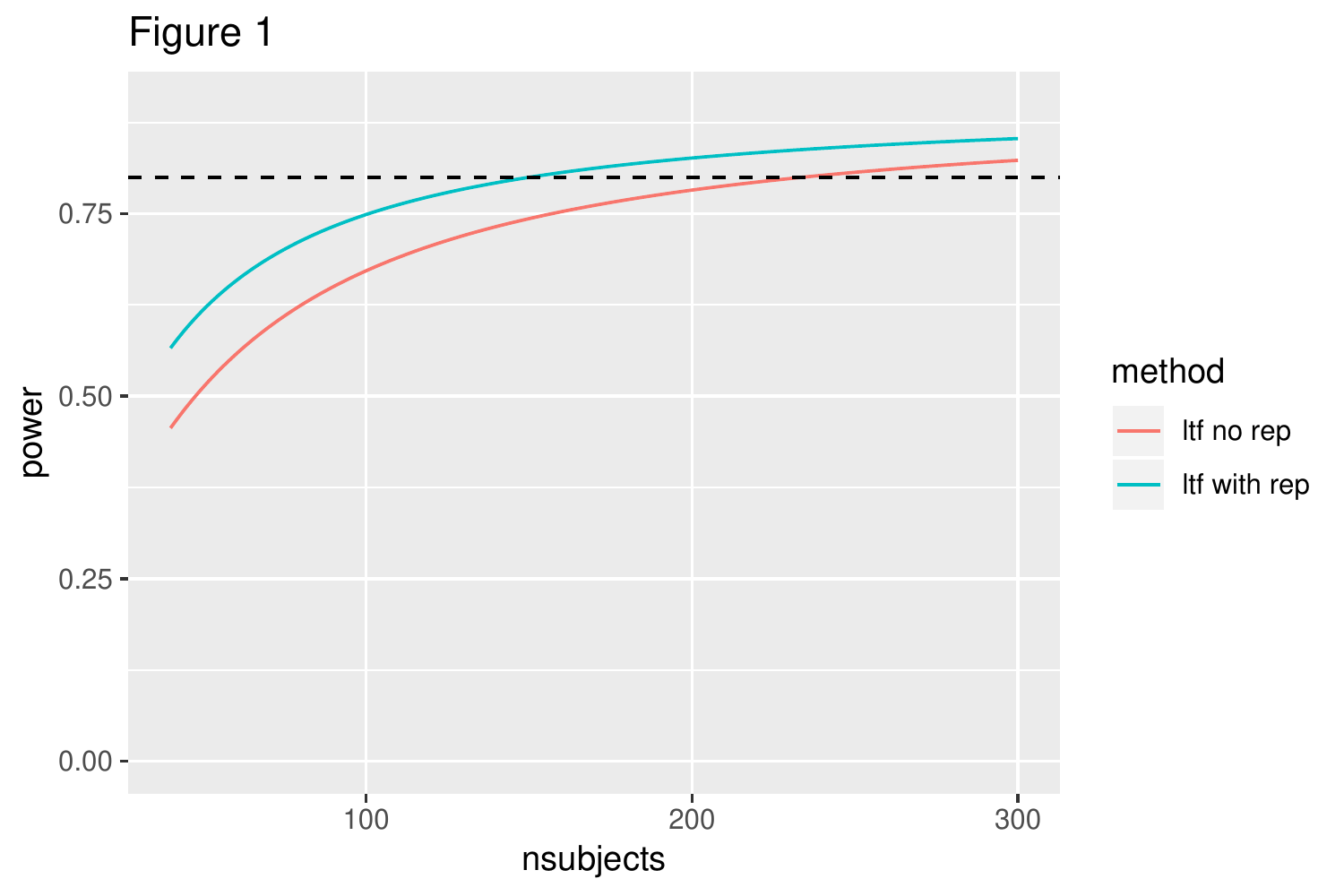} \end{center}

Figure 1 shows the association between the number of subjects per
cluster at baseline and power. The blue line denotes the loss to
follow-up with replacement design (where \(G_i = L_i\) and
\(\rho_S^* = (1 - \bar\lambda)\rho_S\)) and the red line uses the loss
to follow-up with no replacement design (where \(G_i = L_i = 0\) and
\(\rho_S^* = \rho_S - \frac{1}{4}\Big[\frac{\lambda_1}{1-\lambda_1}+\frac{\lambda_2}{1-\lambda_2}\Big]\)).
For the given parameters, the no LTF and complete case designs give
power estimates very similar to the loss to follow-up with replacement
design and so are not plotted here. Figure 1 indicates that for 80\%
power a minimum of 151 subjects per cluster is needed for the loss to
follow-up with replacement design, while a minimum of 235 subjects per
cluster is needed for the loss to follow-up with no replacement design.

\hypertarget{comparisons-of-did-estimators}{%
\section{4. Comparisons of DID
Estimators}\label{comparisons-of-did-estimators}}

In this section we compare versions of the DID estimators. Section 4.1
presents relative efficiency plots. Section 4.2 presents figures
comparing variance as a function of subject autocorrelation \(\rho_S\),
as well as indicating which DID design (reduced cohort or LTF with no
replacement) yields a smaller variance for given follow-up rates.

\hypertarget{comparisons-of-did-estimators-1}{%
\subsection{4.1 Comparisons of DID
Estimators}\label{comparisons-of-did-estimators-1}}

Figure 2 presents relative efficiency plots for three comparisons. LTF
rates for group 1 and group 2 are plotted on the \(x\)- and \(y\)-axes,
while relative efficiency is plotted on the \(z\)-axis as color. The
variances and correlation coefficients used in White represents a
relative efficiency of 1, while more intense color represents larger
relative efficiencies. For all plots, the number of clusters per arm was
30, the number of subjects per cluster was 100 while \(\sigma_C^2\),
\(\sigma_{CT}^2\), \(\sigma_S^2\), and \(\sigma_{ST}^2\) were 0.015,
0.035, 0.76, and 0.19, respectively. The variance values correspond to
an ICC \(\rho\) of 0.05, a cluster autocorrelation \(\rho_C\) of 0.3,
and a subject autocorrelation \(\rho_S\) of 0.8.

\begin{center}\includegraphics{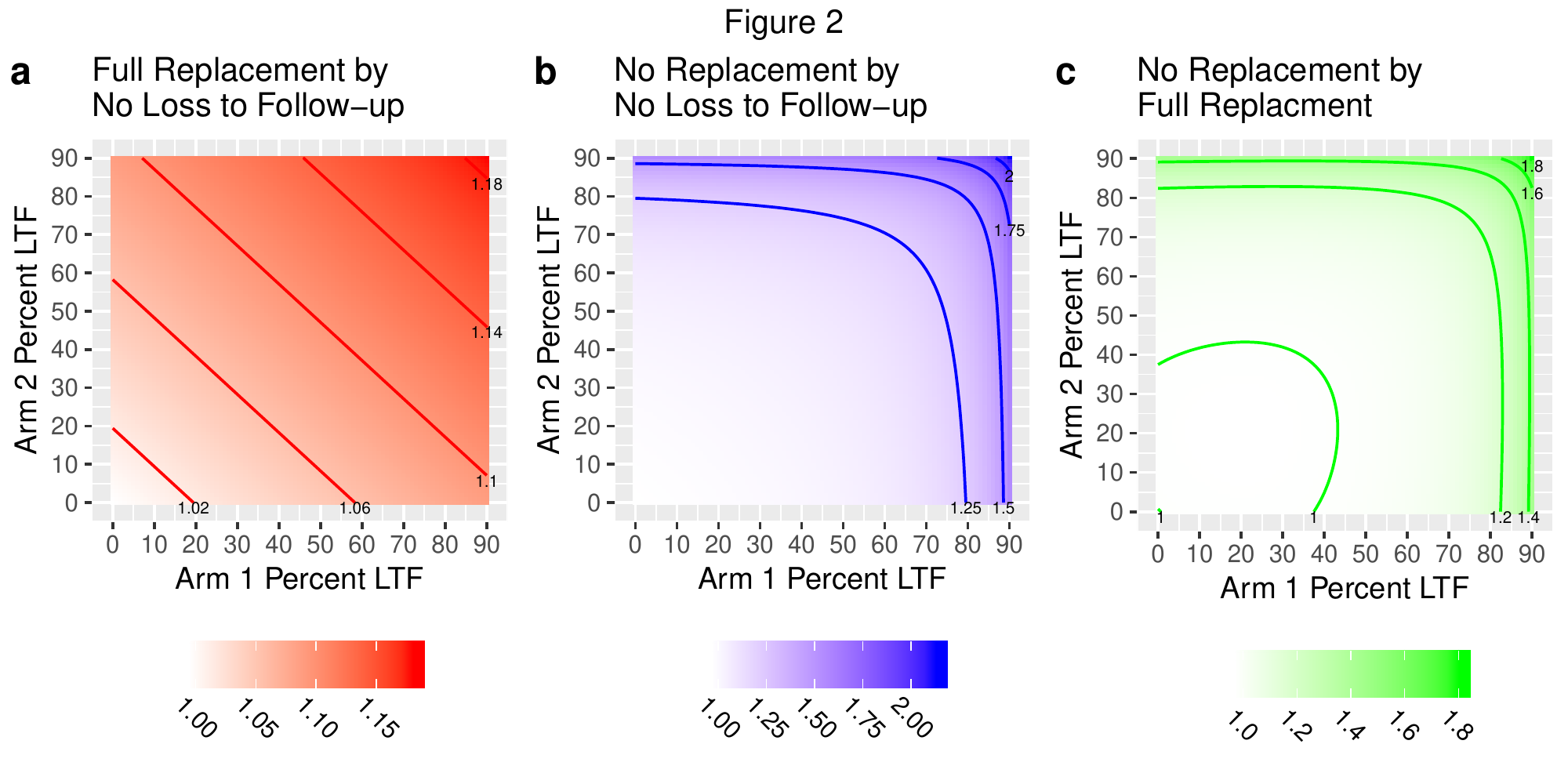} \end{center}

Figure 2a compares the LTF with full replacement variance (where
\(G_i=L_i\) and using \(\rho_S^* = (1 - \bar\lambda)\rho_S\) in equation
\ref{eq:vardid}) to that of the no LTF cohort design
(\(L_i = G_i = 0\)). Figure 2b compares the LTF with no replacement
variance (with \(L_i > G_i = 0\) and using
\(\rho_S^* = \rho_S - \frac{1}{4}\Big[\frac{\lambda_1}{1-\lambda_1}+\frac{\lambda_2}{1-\lambda_2}\Big]\)
in equation \ref{eq:vardid}) to that of the no LTF cohort design. Figure
2c compares the no replacement variance (\(L_i > G_i = 0\)) to the full
replacement variance (\(L_i = G_i\)). In figures 2a and 2b, as percent
lost to follow-up in both arms increases, the variances of the complete
replacement and no replacement estimators are inflated relative to the
no LTF cohort design. In both figures 2a and 2b, when percent LTF in
both groups is low the variance inflation is relatively mild. For
example, if 10\% of units are lost to follow-up in both treatment
groups, figure 2a indicates that the variance of the LTF with full
replacement estimator is only about 1.02 times greater than that of no
LTF cohort design. Similarly, with 10\% lost to follow-up in each arm
figure 2b indicates the variance of the LTF with no replacement
estimator is close to 1. Indeed, at even fairly high loss to follow-up
percentages the inflation factor for the full replacement variance is
low. For example, if the percent lost to follow-up is 80\%, figure 2a
indicates that the LTF with replacement estimator variance is only about
1.035 times that of the no LTF cohort design. However, figure 2b
indicates that, with an 80\% loss to follow-up in both treatment groups,
the variance of the LTF with no replacement estimator is approximately
1.2 times that of the no LTF cohort. Given that the inflation factors
depicted in figure 2a remain fairly close to 1, it's not surprising that
figure 2c bears a stronger resemblance to figure 2b.

In figure 2c, the lower left region of the plot indicates that the full
replacement variance is slightly less than the the no replacement
variance. An explanation for this observation is that the larger sample
size of the full replacement scenario does not overcome the added
uncertainty brought about by gaining new individuals at follow-up.

\hypertarget{comparison-of-reduced-cohort-and-ltf-with-no-replacement-designs}{%
\subsection{4.2 Comparison of Reduced Cohort and LTF with no replacement
Designs}\label{comparison-of-reduced-cohort-and-ltf-with-no-replacement-designs}}

A common strategy for power calculation is to assume a cohort design
using the expected cluster size at follow-up, assuming no LTF. In
effect, individuals who were present at baseline but not at follow-up
are discarded in the power calculation. If the data are MCAR this won't
introduce bias, but the smaller sample size will reduce power. The
variance for this calculation is given by equation \ref{eq:vardid},
using the expected follow-up cluster size for \(K\). This will be
referred to as the reduced cohort DID design.

An alternative would be to use the LTF with no replacement formula give
in equation \ref{vardidltfnorep}, which uses the baseline observations
of individuals lost to follow-up. Given that the total sample size in
this equation is greater than the reduced cohort case, it might be
expected to produce a lower variance, and therefore greater power, for a
given effect size. This will be referred to as the LTF with no
replacement DID design.

Here we compare the DID estimator variance of the reduced cohort design
with the variance of the LTF with no replacement design. Figure 3
displays variances plotted against subject autocorrelation for the LTF
no replacment design (100 subjects per cluster at baseline, 50 per
cluster lost to follow-up) and the reduced cohort (50 subjects per
cluster at baseline and follow-up), with treatment and control groups
having 4 and 30 clusters each. Cluster autocorrelation, cluster
variance, and subject variance were set at 0.3, 0.05, and 0.95,
respectively. Despite having a smaller overall sample size, at values of
\(\rho_S\) greater than 0.5 the DID estimator from the smaller reduced
cohort design has less variance than the LTF with no replacement design.

\begin{center}\includegraphics{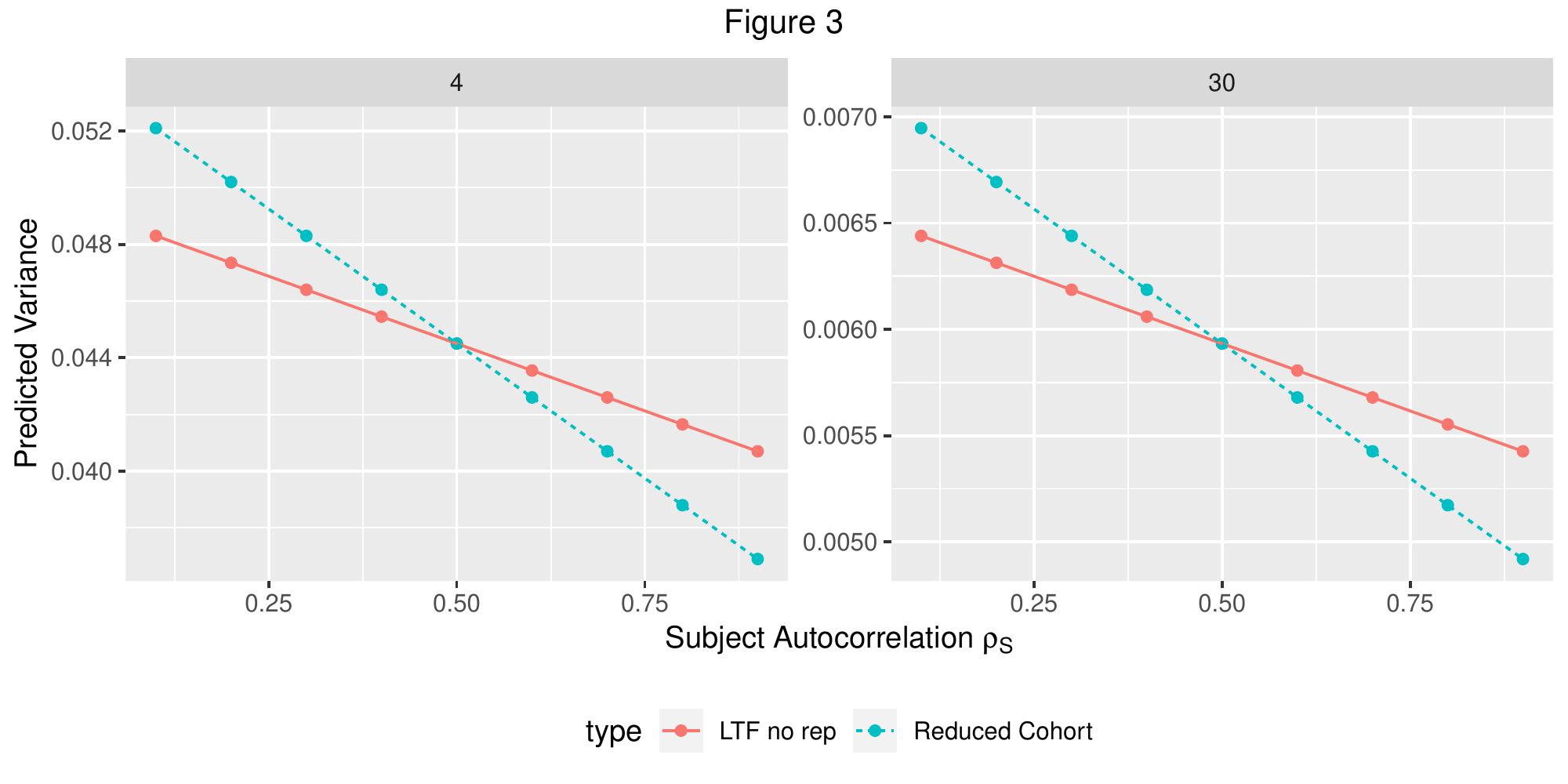} \end{center}

Another scenario of interest is when loss to follow-up is different for
treatment and control arms. Let \(\hat{\beta}_{3,R}\) be the DID
estimator in the reduced cohort scenario. Let \(\hat{\beta}_{3,LTF}\) be
the DID estimator in the LTF with no replacement scenario. Let \(K\) be
the sample size per cluster for both treatment and control at baseline,
so all clusters have the same size at baseline. Let \(K_1\) and \(K_2\)
be the sample sizes per cluster the control and treatment groups at
follow-up, respectively. Suppose the cluster sizes in each arm are
different - without loss of generality, \(K_1 < K_2\). Then the
following inequality

\begin{align} \label{ineq0}
Var(\hat{\beta}_{3,R}) & < Var(\hat{\beta}_{3,LTF}) \nonumber \\
4\frac{\sigma_{CT}^2}{J} + 4\frac{\sigma_{ST}^2}{JK_1} &< 4\frac{\sigma_{CT}^2}{J} + 4\frac{\sigma_{ST}^2}{JK} \nonumber \\
  &+\frac{(\sigma_{S}^2 + \sigma_{ST}^2)}{JK}\Big[\frac{\lambda_1}{1-\lambda_1}+\frac{\lambda_2}{1-\lambda_2}\Big] 
\end{align}

is true when \begin{align} \label{eq:followuprates}
1 - \lambda_1 > \frac{(1-\lambda_2)(3 - 4\rho_S)}{(1-\lambda_2)(2 - 4\rho_S) + 1}
\end{align}

That is, when the follow-up rate in the small group is greater than the
value given on the right hand side of inequality \ref{eq:followuprates},
the variance of the reduced cohort scenario will be less than that given
by the LTF with no replacement scenario.

Figure 4 plots follow-up rates in the smaller group versus follow-up
rates in the larger group for several values of \(\rho_S\). The dashed
diagonal line is the identity line and correpsonds to a \(\rho_S\) of
0.50. Values of \(\rho_S\) below 0.50 and above 0.75 are not shown as
they produce follow-up rates in the smaller group that either less than
0 or are greater than those in the larger group. The dot at (0.95, 0.84)
represents the follow-up rates for the two arms in the trial described
by Miller et al (2012) and is discussed further below.

\begin{center}\includegraphics{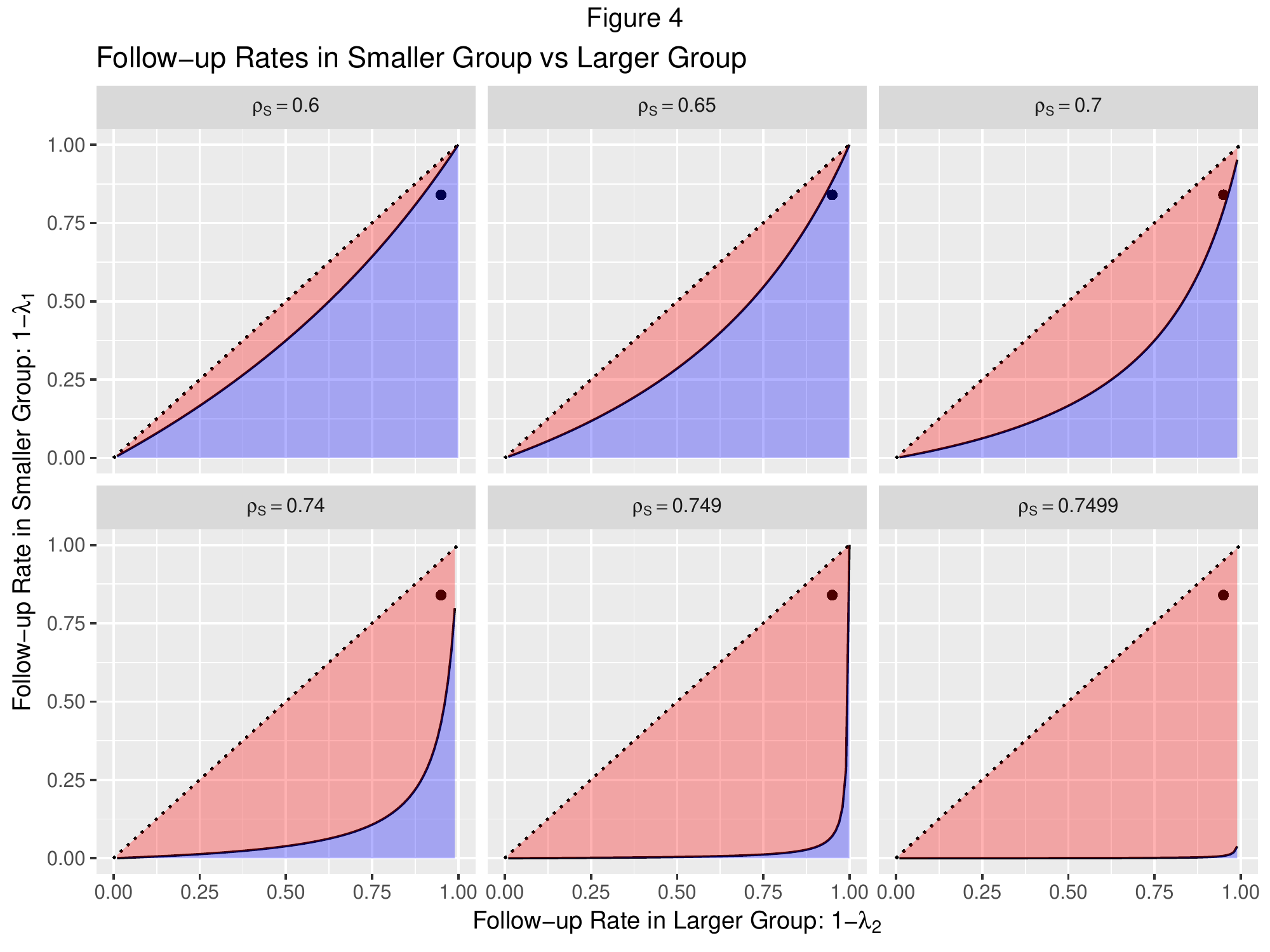} \end{center}

Combinations of follow-up rates in this red region (above the solid
line) indicate that variance for the reduced cohort DID estimator is
less than the variance in the LTF with no replacement DID estimator.
Therefore, sample size calculations for combinations in this region
should use the variance given in equation \ref{eq:vardid}, using the
minimum of treatment or control follow-up cluster size for \(K\). In
addition, analyses should omit those lost to follow-up.

The blue shaded regions correspond to combinations of follow-up rates
and \(\rho_S\) that don't satisfy inequality \ref{eq:followuprates}. In
these regions, the variance of the reduced cohort DID estimator will be
greater than that of the LTF with no replacement DID estimator should be
used. These plots serve as a useful tool for selecting a method to
calculating variance and determining which approach to use in order to
minimize participant risk while still attaining the desired power.

For example, consider the study by Miller et al (2012) in which the
follow-up rates in the two arms were 95\% and 84\% (represented by the
dot in figure 4) and \(\rho_S\) is slightly less than 0.60. These
settings result in placement in the blue region of the top left plot in
figure 4. This suggests the LTF with no replacement DID estimator has
the smaller variance than that of the reduced cohort DID estimator.
However, suppose \(\rho_S\) was 0.70 but the follow-up rates were the
same. In this situation, placement would now be in the red region in the
top right plot in figure 4. This indicates that reduced cohort DID
variance will be smaller than the LTF with no replacement DID variance.

Other observations from figure 4 include the following. For values of
\(\rho_S < 0.50\), the blue region occupies everything under the line of
identity and the LTF with no replacement approach will yield the
smallest variance for any combination of follow-up rates. If
\(\rho_S > 0.5\) and the follow-up rates per cluster between the two
groups are the same, then the line of identity will always be in the red
region and the reduced cohort calculation should be used. For values of
\(\rho_S > 0.75\), the region under the line of identity will be
completely red and the reduced cohort variance computation should be
used.

\hypertarget{conclusion}{%
\section{5. Conclusion}\label{conclusion}}

Relative to the no loss to follow-up (LTF) cohort design, the LTF with
replacement scenario results in mild variance inflation for even
relatively high LTF rates, as shown in Figure 2. In situations where it
is costly or difficult to maintain a cohort but replacements from within
clusters can be found, the LTF with replacement mixture design appears
to be an acceptable approach. During study design, when finding a needed
sample size for the LTF with replacement scenario, one can estimate the
variance with equation \ref{eq:vardid} by multiplying the expected
subject autocorrelation \(\rho_S\) by the expected mean follow-up rate
as shown in equation \ref{vardidwrep}.

The LTF with no replacement scenario also results in only mild variance
inflation relative to no LTF cohort design if LTF rates are low in both
control and treatment arms. However, high LTF rates in at least one of
the arms in the LTF with no replacement scenario results in substantial
variance inflation. Interestingly, the LTF with no replacement design is
more efficient than the LTF with full replacement design when the loss
to follow-up rates are very low. The larger sample size in the full
replacement scenario does not overcome the added uncertainty brought
about by obtaining new individuals at follow-up.

When comparing LTF with no replacement and reduced cohort designs, low
values of \(\rho_S\) indicate that the former has a smaller predicted
variance while large values of \(\rho_S\) indicate the opposite is true,
as shown in Figure 3. Typically, there is no a priori expectation of
different loss to follow-up in each treatment group. Thus, when
\(\rho_S\) is expected to be below 0.5, the LTF with no replacement DID
approach is recommended. One can then use equation \ref{vardidltfnorep}
to compute the variance of the DID estimator. Otherwise, if \(\rho_S\)
is expected to be greater than 0.5, the reduced cohort design is
preferred.

Our results also show that, in general, the LTF with no replacement
scenario outperforms the reduced cohort case when the subject
autocorrelation \(\rho_S\) is low and loss to follow-up rates are high,
as shown in Figure 4. However, when \(\rho_S\) is large and LTF rates
are low, the reduced cohort design yields a smaller variance. In these
situations, the strong subject autocorrelation provides more information
through the DID estimator than that provided from the means at baseline
and follow-up. In practice, with estimates of \(\rho_S\) and the
follow-up rates in each arm, Figure 4 can be used to determine which
design - LTF with no replacement or reduced cohort - is preferred over
the other. Follow-up rates and \(\rho_S\) values resulting in placement
in the red region indicate that the reduced cohort design is preferred,
while placement in the blue region indicates that the LTF with no
replacement design is preferred.

A limitation of the methods discussed above is that the variance
parameters \(\sigma_C^2\), \(\sigma_{CT}^2\), \(\sigma_S^2\), and
\(\sigma_{ST}^2\) are rarely all reported in study results. Knowledge of
these four parameters is needed for finding the ICC, \(\rho_C\), and
\(\rho_S\) parameters. In addition, not all cohort designs may have
replacement subjects available, limitind consideration to the no LTF and
reduced cohort settings.

In summary, our results indicate that if individuals are available to
replace those lost to follow-up, the LTF with replacement design
performs similarly to the no LTF design. An exception is when the loss
to follow-up rates in each arm are relatively low. In this case the LTF
with no replacement design has a smaller variance. If it is not possible
to replace those lost to follow-up, then the choice of using the LTF
with no replacement and reduced cohort designs depends largely on the
magnitude of subject autocorrelation \(\rho_S\). If \(\rho_S\) is
expected to be below 0.50, the LTF with no replacement design should be
used. If \(\rho_S\) is larger than 0.50 and the follow-up rates in the
arms are expected to be very different, then the LTF with no replacement
design is likely to be preferred. However, if \(\rho_S\) is large and
the follow-up rates in each arm are expected to be the same or similar,
the reduced cohort design is preferred. Finally, if \(\rho_S\) is
expected to be very large (greater than 0.75), then the reduced cohort
design should be used regardless of the follow-up rates in each arm.

\newpage

\hypertarget{appendix}{%
\section{Appendix}\label{appendix}}

\hypertarget{a.1-did-estimator}{%
\subsection{A.1 DID Estimator}\label{a.1-did-estimator}}

Let \(J\) be the number of clusters per arm. Let \(K_{itj}\) be the
number of subjects in cluster \(ij\) at time \(t\). Define the mean of
arm \(i\) (\(i = 1,2\)) at time \(t\) (\(t = 1,2\)) as:
\begin{equation} \label{eq:yitbar}
\bar{y}_{it} = \sum_{j=1}^J \sum_{k = 1}^{K_{itj}} \frac{y_{itjk}}{JK_{itj}}
\end{equation}

and \(\delta_{11} = 1\), \(\delta_{12} = -1\), \(\delta_{21} = -1\), and
\(\delta_{22} = 1\). Then the DID estimator \(\hat\beta_3\) is given by
\begin{align} \label{eq:d0}
\hat\beta_3 =& \sum_{i=1}^2\sum_{t = 1}^2\delta_{it}\bar{y}_{it} \nonumber \\
  =& (\bar{y}_{22} - \bar{y}_{21}) - (\bar{y}_{12} - \bar{y}_{11}) \nonumber \\
  =& \bar{y}_{11} - \bar{y}_{12} - \bar{y}_{21} + \bar{y}_{22} \nonumber \\
  =& \sum_{j=1}^J \sum_{k = 1}^{K_{11j}} \frac{y_{11jk}}{JK_{11j}} - \sum_{j=1}^J \sum_{k = 1}^{K_{12j}} \frac{y_{12jk}}{JK_{12j}} - \sum_{j=1}^J \sum_{k = 1}^{K_{21j}} \frac{y_{21jk}}{JK_{21j}} + \sum_{j=1}^J \sum_{k = 1}^{K_{22j}} \frac{y_{22jk}}{JK_{22j}}
\end{align}

The variance of the DID estimator is: \begin{align} \label{var.beta3}
\text{Var}(\hat\beta_3) &= \text{\text{Var}}(\bar{y}_{11} - \bar{y}_{12} - \bar{y}_{21} + \bar{y}_{22}) \nonumber \\
        &= \text{Var}(\bar{y}_{11}) + \text{Var}(\bar{y}_{12}) + \text{Var}(\bar{y}_{21}) + \text{Var}(\bar{y}_{22}) - 2\text{Cov}(\bar{y}_{11},\bar{y}_{12}) - 2\text{Cov}(\bar{y}_{21},\bar{y}_{22})
\end{align}

because for \(i \ne i^\prime\),
Cov(\(\bar{y}_{it},\bar{y}_{i^\prime t^\prime}\)) = 0.

The cluster variance of the mean response for arm \(i\) at time \(t\),
\(\text{Var}(\bar{y}_{it})\), is given by: \begin{align} \label{var.yit}
\text{Var}(\bar{y}_{it}) &= \text{Var}\Bigg(\sum_{j=1}^J \sum_{k = 1}^{K_{itj}} \frac{y_{itjk}}{JK_{itj}} \Bigg) \nonumber \\
                         &= \frac{1}{J^2}\Big[\text{Var}\Big(\sum_{k = 1}^{K_{it1}} \frac{y_{it1k}}{K_{it1}}\Big) + \text{Var}\Big(\sum_{k = 1}^{K_{it2}} \frac{y_{it2k}}{K_{it2}}\Big) + \dots + \text{Var}\Big(\sum_{k = 1}^{K_{itJ}} \frac{y_{itJk}}{K_{itJ}} \Big) \Big] 
\end{align}

because for \(j \ne j^\prime\), Cov(\(y_{itjk},y_{itj^\prime k}\)) = 0.

For arm \(i\), the covariance between mean response at baseline and
follow-up is given by \begin{align} \label{cov.y12}
\text{Cov}(\bar{y}_{i1},\bar{y}_{i2}) &= \text{Cov}\Bigg(\sum_{j=1}^J \sum_{k = 1}^{K_{i1j}} \frac{y_{i1jk}}{JK_{i1j}},\sum_{j=1}^J \sum_{k = 1}^{K_{i2j}} \frac{y_{i2jk}}{JK_{i2j}} \Bigg) \nonumber \\
                                      &= \frac{1}{J^2}\sum_{j=1}^J\sum_{k=1}^{K_{i1j}}\sum_{k^\prime=1}^{K_{i2j}} \text{Cov}(y_{i1jk},y_{i2jk^\prime}) 
\end{align}

The expressions given in equations \ref{var.yit} and \ref{cov.y12} do
not make assumptions about cluster sizes. In subsequent sections we will
make simplifying assumptions about the cluster sizes.

\hypertarget{a.2-did-estimator-with-modifications-for-loss-to-follow-up-and-replacement}{%
\subsection{A.2 DID Estimator with modifications for loss to follow-up
and
replacement}\label{a.2-did-estimator-with-modifications-for-loss-to-follow-up-and-replacement}}

In this section we derive the variance for the loss to follow-up (LTF)
with partial repalcement scenario with variable LTF for each cluster. In
this setting the \(ij\) cluster at baseline contains \(K\) individuals,
loses \(L_{ij}\) individuals at follow-up, and partially replaces
\(G_{ij}\) individuals at follow-up. A matrix of the \(ij\) cluster is
presented below (the \(ij\) subscripts are omitted for clarity).

\begin{center}
\begin{tikzpicture}
\matrix [matrix of math nodes,nodes in empty cells,left delimiter={[},right delimiter={]},
nodes={anchor=south}
] (m) {
\sigma^2 &          &        &          &          &          &          &        &          &          &          &          &          &          &          &          & \\
         & \sigma^2 &        &          &          &          &          &        &          &          &          &\sigma_C^2&          &          &          &\sigma_C^2& \\
         &          &        &          &          &          &          &        &          &          &          &          &          &          &          &          & \\
         &          &        & \sigma^2 &          &          &          &        &          &          &          &          &          &          &          &          & \\
         &          &        &          & \sigma^2 &          &          &        &          &          &          &          &          &          &          &          & \\
         &          &        &          &          &          &          &        &          &          &          &          &          &          &          &          & \\
         &          &        &          &          &          &          &        &          &          &          &          &          &          &          &\sigma_C^2& \\
         &          &        &          &          &          &          &        &          &          &          &          &          &          &          &          & \\
         &          &        &          &          &          &          &        & \sigma^2 &          &          &          &          &          &          &          & \\
         &          &        &          &          &          &          &        &          & \sigma^2 &          &          &          &          &          &          & \\
         &          &        &          &          &          &          &        &          &          &          &          &          &          &          &          & \\
         &          &        &          &          &          &          &        &          &          &          &          &          &          &          &          & \\
         &          &        &          &          &          &          &        &          &          &          &          &          &          &          &          & \\
         &          &        &          &          &          &          &        &          &          &          &          &          & \sigma^2 &          &          & \\
         &          &        &          &          &          &          &        &          &          &          &          &          &          & \sigma^2 &          & \\
         &          &        &          &          &          &\sigma_C^2&        &          &          &          &          &          &          &          &          & \\
         &          &        &          &          &          &          &        &          &          &          &          &          &          &          &          & \sigma^2 \\};
         
\draw[very thick,opacity=0.3] (m-1-1.north west) rectangle (m-9-9.south east);
\draw[very thick,opacity=0.3] (m-10-10.north west) rectangle (m-17-17.south east);
         
\draw[dashed] ($0.5*(m-1-4.north west)+0.5*(m-1-5.north east)$) -- ($0.5*(m-17-4.south west)+0.5*(m-17-5.south east)$);
\draw[dashed] ($0.5*(m-1-14.north west)+0.5*(m-1-15.north east)$) -- ($0.5*(m-17-14.south west)+0.5*(m-17-15.south east)$);

\draw[dashed] (m-4-1.south west) -- (m-4-17.south east);
\draw[dashed] (m-14-1.south west) -- (m-14-17.south east);

\node[above=12pt of m-1-1] (top-1) {};
\node[above=12pt of m-1-4] (top-4) {};
\node[above=12pt of m-1-5] (top-5) {};
\node[above=12pt of m-1-9] (top-9) {};
\node[above=12pt of m-1-10] (top-10) {};
\node[above=12pt of m-1-14] (top-14) {};
\node[above=12pt of m-1-15] (top-15) {};
\node[above=12pt of m-1-17] (top-17) {};

\node[left=12pt of m-1-1] (left-1) {};
\node[left=12pt of m-4-1] (left-4) {};
\node[left=12pt of m-5-1] (left-5) {};
\node[left=12pt of m-9-1] (left-9) {};
\node[left=12pt of m-10-1] (left-10) {};
\node[left=12pt of m-14-1] (left-14) {};
\node[left=12pt of m-15-1] (left-15) {};
\node[left=12pt of m-17-1] (left-17) {};

\node[fit=(m-2-5)(m-2-9)]{$\sigma_C^2 + \sigma_{CT}^2$};

\node[fit=(m-7-1)(m-7-4)]{$\sigma_C^2 + \sigma_{CT}^2$};
\node[fit=(m-7-11)(m-7-13)]{$\sigma_C^2 + \sigma_S^2$};

\node[fit=(m-12-2)(m-12-3)]{$\sigma_C^2$};
\node[fit=(m-12-6)(m-12-8)]{$\sigma_C^2 + \sigma_S^2$};
\node[fit=(m-12-15)(m-12-17)]{$\sigma_C^2 + \sigma_{CT}^2$};

\node[fit=(m-16-2)(m-16-3)]{$\sigma_C^2$};
\node[fit=(m-16-11)(m-16-13)]{$\sigma_C^2 + \sigma_S^2$};

\node[yshift=-0.25cm,rectangle,above delimiter=\{] (del-top-1) at ($0.5*(top-1.south) + 0.5*(top-4.south)$) {\tikz{\path (top-1.south west) rectangle (top-4.north east);}};
\node[above=10pt] at (del-top-1.north) {$L$};
\node[yshift=-0.25cm,rectangle,above delimiter=\{] (del-top-2) at ($0.5*(top-5.south) + 0.5*(top-9.south)$) {\tikz{\path (top-5.south west) rectangle (top-9.north east);}};
\node[above=10pt] at (del-top-2.north) {$K-L$};
\node[yshift=-0.25cm,rectangle,above delimiter=\{] (del-top-3) at ($0.5*(top-10.south) + 0.5*(top-14.south)$) {\tikz{\path (top-10.south west) rectangle (top-14.north east);}};
\node[above=10pt] at (del-top-3.north) {$K-L$};
\node[yshift=-0.25cm,rectangle,above delimiter=\{] (del-top-4) at ($0.5*(top-15.south) + 0.5*(top-17.south)$) {\tikz{\path (top-15.south west) rectangle (top-17.north east);}};
\node[above=10pt] at (del-top-4.north) {$G$};

\node[rectangle,left delimiter=\{] (del-left-1) at ($0.5*(left-1.east) + 0.5*(left-4.east)$) {\tikz{\path (left-1.north east) rectangle (left-4.south west);}};
\node[left=10pt] at (del-left-1.west) {$L$};
\node[rectangle,left delimiter=\{] (del-left-2) at ($0.5*(left-5.east) + 0.5*(left-9.east)$) {\tikz{\path (left-5.north east) rectangle (left-9.south west);}};
\node[left=10pt] at (del-left-2.west) {$K-L$};
\node[rectangle,left delimiter=\{] (del-left-3) at ($0.5*(left-10.east) + 0.5*(left-14.east)$) {\tikz{\path (left-10.north east) rectangle (left-14.south west);}};
\node[left=10pt] at (del-left-3.west) {$K-L$};
\node[rectangle,left delimiter=\{] (del-left-4) at ($0.5*(left-15.east) + 0.5*(left-17.east)$) {\tikz{\path (left-15.north east) rectangle (left-17.south west);}};
\node[left=10pt] at (del-left-4.west) {$G$};

\end{tikzpicture}
\end{center}

where
\(\sigma^2 = \sigma_C^2 + \sigma_{CT}^2 + \sigma_S^2 + \sigma_{ST}^2\).
The upper left and lower right quadrants of this matrix (enclosed in
light gray boxes) represent cluster covariances at baseline and
follow-up, respectively. In the area enclosed by the light gray boxes,
all off-diagonal elements are \(\sigma_C^2 + \sigma_{CT}^2\). The
off-diagonal blocks represent covariances from baseline to follow-up.
The covariance parameter indicated within a submatrix demarcated by
dashed lines implies that all elements in that submatrix are that
covariance value.

The table below shows the covariances for cluster \(ij\) at time \(t\).

\footnotesize
\begin{adjustwidth}{-0.5cm}{-0.5cm}
\begin{center}
\begin{tabular}{lcccc}
\toprule
        & \multicolumn{2}{c}{Observations} & \multicolumn{1}{c}{Covariance} & \multicolumn{1}{c}{Sum} \\
\midrule
$t = 1$ & $y_{i1jk}$ & $y_{i1j1}$ & $\sigma_{C}^2 + \sigma_{CT}^2$ & \multirow{6}{*}{$K\sigma_{C}^2 + K\sigma_{CT}^2 + \sigma_{S}^2 + \sigma_{ST}^2$} \\
        & $y_{i1jk}$ & $y_{i1j2}$ & $\sigma_{C}^2 + \sigma_{CT}^2$ & \\
        & \vdots     & \vdots     & \multicolumn{1}{c}{\vdots}     & \\
        & $y_{i1jk}$ & $y_{i1jk}$ & $\sigma_{C}^2 + \sigma_{CT}^2 + \sigma_{S}^2 + \sigma_{ST}^2$ & \\
        & \vdots     & \vdots     & \multicolumn{1}{c}{\vdots}     & \\
        & $y_{i1jk}$ & $y_{i1jK}$ & $\sigma_{C}^2 + \sigma_{CT}^2$ & \\
\midrule
$t = 2$, & $y_{i2jk}$ & $y_{i2j(L_{ij}+1)}$ & $\sigma_{C}^2 + \sigma_{CT}^2$ & \multirow{6}{*}{$(K-L_{ij}+G_{ij})\sigma_{C}^2 + (K-L_{ij}+G_{ij})\sigma_{CT}^2 + \sigma_{S}^2 + \sigma_{ST}^2$} \\
$L_{ij} < k \le K-L_{ij}$  & $y_{i2jk}$ & $y_{i2j(L_{ij}+2)}$ & $\sigma_{C}^2 + \sigma_{CT}^2$ & \\
        & \vdots     & \vdots     & \multicolumn{1}{c}{\vdots}     & \\
        & $y_{i2jk}$ & $y_{i2jk}$ & $\sigma_{C}^2 + \sigma_{CT}^2 + \sigma_{S}^2 + \sigma_{ST}^2$ & \\
        & \vdots     & \vdots     & \multicolumn{1}{c}{\vdots}     & \\
        & $y_{i2jk}$ & $y_{i2j(K-L_{ij} + G_{ij})}$ & $\sigma_{C}^2 + \sigma_{CT}^2$ & \\
\midrule
$t = 2$, & $y_{i2jk}$ & $y_{i2j(L_{ij}+1)}$ & $\sigma_{C}^2 + \sigma_{CT}^2$ & \multirow{6}{*}{$(K-L_{ij}+G_{ij})\sigma_{C}^2 + (K-L_{ij}+G_{ij})\sigma_{CT}^2$} \\
$k > K-L_{ij}$        & $y_{i2jk}$ & $y_{i2j(L_{ij}+2)}$ & $\sigma_{C}^2 + \sigma_{CT}^2$ & \\
        & \vdots     & \vdots     & \multicolumn{1}{c}{\vdots}     & \\
        & $y_{i2jk}$ & $y_{i2j(K-L_{ij} + G_{ij})}$ & $\sigma_{C}^2 + \sigma_{CT}^2$ & \\
\bottomrule
\end{tabular}
\end{center}
\end{adjustwidth}
\normalsize

At baseline, there are \(K\) blocks, so for cluster \(j\) the sum of the
covariances is:

\begin{equation*}
K(K\sigma_{C}^2 + K\sigma_{CT}^2 + \sigma_{S}^2 + \sigma_{ST}^2)
\end{equation*}

Thus equation \ref{var.yit} becomes:

\begin{align} \label{eq:vardit1}
\text{Var}(\bar{y}_{it}) &= \frac{1}{(JK)^2}\Big[\text{Var}\Big(\sum_{k=1}^K y_{it1k}\Big) + \text{Var}\Big(\sum_{k=1}^K y_{it2k}\Big) + \dots + \text{Var}\Big(\sum_{k=1}^K y_{itJk} \Big) \Big] \nonumber \\
                         &= \frac{1}{(JK)^2}\Big[JK(K\sigma_{C}^2 + K\sigma_{CT}^2 + \sigma_{S}^2 + \sigma_{ST}^2)\Big] \nonumber \\
                         &= \frac{1}{JK}(K\sigma_{C}^2 + K\sigma_{CT}^2 + \sigma_{S}^2 + \sigma_{ST}^2) 
\end{align}

For clusters at follow-up, the result is similar, save that the cluster
size at follow-up is \(K-L_{ij}+G_{ij}\). So

\begin{align} \label{eq:vardit1z2}
\text{Var}(\bar{y}_{i2}) &= \frac{1}{J^2}\Bigg[
\text{Var}\Bigg(\frac{\sum_{k=1}^{K-L_{i1}+G_{i1}} y_{i21k}}{K - L_{i1}+G_{i1}}\Bigg) + 
\text{Var}\Bigg(\frac{\sum_{k=1}^{K-L_{i2}+G_{i2}} y_{i22k}}{K - L_{i2}+G_{i2}}\Bigg) + \dots + 
\text{Var}\Bigg(\frac{\sum_{k=1}^{K-L_{iJ}+G_{iJ}} y_{i2Jk}}{K - L_{iJ}+G_{iJ}}\Bigg) \Bigg] \nonumber \\
                         &= \frac{1}{J^2}
\Bigg[
\frac{\text{Var}\Big(\sum_{k=1}^{K-L_{i1}+G_{i1}} y_{i21k}\Big)}{(K - L_{i1}+G_{i1})^2} + 
\frac{\text{Var}\Big(\sum_{k=1}^{K-L_{i2}+G_{i2}} y_{i22k}\Big)}{(K - L_{i2}+G_{i2})^2} + \dots + 
\frac{\text{Var}\Big(\sum_{k=1}^{K-L_{iJ}+G_{iJ}} y_{i2Jk}\Big)}{(K - L_{iJ}+G_{iJ})^2} \Bigg] 
\end{align}

Because the denominators are not all the same, equation
\ref{eq:vardit1z2} cannot be simplified further. We need to make
assumptions about the \(L_{ij}\)'s to proceed. In the preceding section
we assumed \(L_{ij} = L_{i^\prime j^\prime}\) and
\(G_{ij} = G_{i^\prime j^\prime}\) for any \(i\) and \(j\). This means
that the losses to follow-up for each cluster are the same and the
numbers replaced at follow-up for each cluster are the same, but the
number of replacments does not necessary equal the number lost. Here we
assume that the loss to follow-up and replacement at follow-up varies by
treatment arm. Thus for arm \(i\) we have

\begin{equation} \label{eq:varit3a2}
\text{Var}(\bar{y}_{i2}) = \frac{1}{J(K-L_{i}+G_{i})}[(K-L_{i}+G_{i})\sigma_{C}^2 + (K-L_{i}+G_{i})\sigma_{CT}^2 + \sigma_{S}^2 + \sigma_{ST}^2] 
\end{equation}

where \(L_i\) and \(G_i\) are the number lost to follow-up and replaced
at follow-up, respectively, in arm \(i\).

The covariances for observations in cluster \(ij\) across time are given
in the following tables:

\begin{center}
\begin{tabular}{lcccc}
\toprule
 & $y_{i1jk}$ & $y_{i2jk}$ & \multicolumn{1}{c}{Covariance} & \multicolumn{1}{c}{Sum in Block} \\
\midrule
$k \le L_i$ & $y_{i1jk}$ & $y_{i2j1}$ & 0 & \multirow{7}{*}{$(K-L_i+G_i)\sigma_{C}^2$} \\
            & \vdots & \vdots & \multicolumn{1}{c}{\vdots} & \\
            & $y_{i1jk}$ & $y_{i2jk}$ & 0 & \\
            & \vdots & \vdots & \multicolumn{1}{c}{\vdots} & \\
            & $y_{i1jk}$ & $y_{i2jL_i}$ & 0 & \\
            \cmidrule{2-4}
            & $y_{i1jk}$ & $y_{i2j(L_i+1)}$ & $\sigma_{C}^2$ & \\
            & \vdots     & \vdots     & \multicolumn{1}{c}{\vdots}     & \\
            & $y_{i1jk}$ & $y_{i2j(K-L_i)}$ & $\sigma_{C}^2$ & \\
            \cmidrule{2-4}
            & $y_{i1jk}$ & $y_{i2j(K-L_i+1)}$ & $\sigma_{C}^2$ & \\
            & \vdots     & \vdots     & \multicolumn{1}{c}{\vdots}     & \\
            & $y_{i1jk}$ & $y_{i2j(K-L_i+G_i)}$ & $\sigma_{C}^2$ & \\
\midrule[.5pt]
$L_i < k \le K-L_{i}$ & $y_{i1jk}$ & $y_{i2j1}$ & 0 & \multirow{9}{*}{$(K-L_i+G_i)\sigma_{C}^2 + \sigma_{S}^2$} \\
            & \vdots & \vdots & \multicolumn{1}{c}{\vdots} & \\
            & $y_{i1jk}$ & $y_{i2jL_i}$ & 0 & \\
            \cmidrule{2-4}
            & $y_{i1jk}$ & $y_{i2j(L_i+1)}$ & $\sigma_{C}^2$ & \\
            & \vdots & \vdots & \multicolumn{1}{c}{\vdots} & \\
            & $y_{i1jk}$ & $y_{i2jk}$ & $\sigma_{C}^2 + \sigma_{S}^2$ & \\
            & \vdots & \vdots & \multicolumn{1}{c}{\vdots} & \\
            & $y_{i1jk}$ & $y_{i2j(K-L_i)}$ & $\sigma_{C}^2$ & \\
            \cmidrule{2-4}
            & $y_{i1jk}$ & $y_{i2j(K-L_i+1)}$ & $\sigma_{C}^2$ & \\
            & \vdots & \vdots & \multicolumn{1}{c}{\vdots} & \\
            & $y_{i1jk}$ & $y_{i2j(K-L_i+G_i)}$ & $\sigma_{C}^2$ & \\
\bottomrule
\end{tabular}
\end{center}

In the \(k > K-L_i\) block of the preceding table, the ``Sum in Block''
is \((K-L_i)\sigma_{C}^2\) instead of \(K\sigma_{C}^2\) to avoid double
counting cases in the intersection with the \(k \le L_i\) block.

For \(k \le L_i\), the sum of the \(L_i\) covariances is
\(L_i(K-L_i+G_i)\sigma_{C}^2\). For \(L_i < k \le K-L_{i}\), the sum of
the \(K-L_i\) covariances is
\((K-L_i)[(K-L_i+G_i)\sigma_{C}^2+\sigma_{S}^2]\). Thus the sum of the
covariances for cluster \(ij\) across time is:

\footnotesize
\begin{align*}
    L_i(K-L_i+G_i)\sigma_{C}^2 + (K-L_i)[(K-L_i+G_i)\sigma_{C}^2+\sigma_{S}^2]
 &= L_i(K-L_i+G_i)\sigma_{C}^2 + K(K-L_i+G_i)\sigma_{C}^2 + K\sigma_{S}^2 - L_i(K-L_i+G_i)\sigma_{C}^2 - L_i\sigma_{S}^2 \nonumber \\
 &= K(K-L_i+G_i)\sigma_{C}^2 + K\sigma_{S}^2 - L_i\sigma_{S}^2 \nonumber \\
 &= K(K-L_i+G_i)\sigma_{C}^2 + (K-L_i)\sigma_{S}^2 \nonumber \\
 &= (K-L_i+G_i)(K\sigma_{C}^2 + \frac{K-L_i}{K-L_i+G_i}\sigma_{S}^2)
\end{align*}
\normalsize

Thus the covariance becomes:

\begin{align} \label{eq:cov0z2}
\text{Cov}(\bar{y}_{i1},\bar{y}_{i2}) &= \text{Cov}\Bigg(\frac{\sum_{j=1}^J\sum_{k=1}^K y_{i1jk}}{JK},\frac{\sum_{j^\prime=1}^J\sum_{k^\prime=1}^{K-L_{i}+G_i} y_{i2j^\prime k^\prime}}{K-L_{i}+G_i} \Bigg) \nonumber \\
                                      &= \frac{1}{J^2K(K-L_i+G_i)}\sum_{j=1}^J\sum_{k=1}^K \sum_{j^\prime=1}^J\sum_{k^\prime=1}^{K-L_i+G_i} \text{Cov}(y_{i1jk},y_{i2j^\prime k^\prime}) \nonumber \\
                                      &= \frac{1}{J^2K(K-L_i+G_i)}\sum_{j=1}^J\sum_{k,k^\prime=1}^{K-L_i+G_i} \text{Cov}(y_{i1jk},y_{i2j k^\prime}) \nonumber \\
                                      &= \frac{1}{J^2K(K-L_i+G_i)}\sum_{j=1}^J(K-L_i+G_i)(K\sigma_{C}^2 + \frac{K-L_i}{K-L_i+G_i}\sigma_{S}^2) \nonumber \\
                                      &= \frac{1}{J^2K(K-L_i+G_i)}J(K-L_i+G_i)(K\sigma_{C}^2 + \frac{K-L_i}{K-L_i+G_i}\sigma_{S}^2) \nonumber \\
                                      &= \frac{1}{JK}(K\sigma_{C}^2 + \frac{K-L_i}{K-L_i+G_i}\sigma_{S}^2)
\end{align}

Using the results from equations \ref{eq:varit3a2} and \ref{eq:cov0z2},
equation \ref{var.beta3} becomes: \small
\begin{align} \label{eq:vard4a2}
\text{Var}(\hat\beta_3) &= \text{Var}(\bar{y}_{11}) + \text{Var}(\bar{y}_{12}) + \text{Var}(\bar{y}_{21}) + \text{Var}(\bar{y}_{22}) - 2\text{Cov}(\bar{y}_{11},\bar{y}_{12}) - 2\text{Cov}(\bar{y}_{21},\bar{y}_{22}) \nonumber \\
        &= \frac{2}{JK}[K\sigma_{C}^2 + K\sigma_{CT}^2 + \sigma_{S}^2 + \sigma_{ST}^2] \nonumber \\
        &+ \frac{1}{J(K-L_1+G_1)}[(K-L_1+G_1)\sigma_{C}^2 + (K-L_1+G_1)\sigma_{CT}^2 + \sigma_{S}^2 + \sigma_{ST}^2] \nonumber \\
        &+ \frac{1}{J(K-L_2+G_2)}[(K-L_2+G_2)\sigma_{C}^2 + (K-L_2+G_2)\sigma_{CT}^2 + \sigma_{S}^2 + \sigma_{ST}^2] \nonumber \\
        &- 2\frac{1}{JK}\Big(K\sigma_{C}^2 + \frac{K-L_1}{K-L_1+G_1}\sigma_{S}^2\Big) - 2\frac{1}{JK}\Big(K\sigma_{C}^2 + \frac{K-L_2}{K-L_2+G_2}\sigma_{S}^2\Big) \nonumber \\
        &= 4\frac{\sigma_{CT}^2}{J} + 2\frac{\sigma_{S}^2}{JK} + 2\frac{\sigma_{ST}^2}{JK} + \frac{\sigma_{S}^2 + \sigma_{ST}^2}{J(K-L_1+G_1)} + \frac{\sigma_{S}^2+\sigma_{ST}^2}{J(K-L_2+G_2)} - 2\frac{\sigma_{S}^2}{JK}\Big(\frac{K-L_1}{K-L_1+G_1}+\frac{K-L_2}{K-L_2+G_2}\Big) 
\end{align} \normalsize

\hypertarget{loss-to-follow-up-with-complete-replacement}{%
\subsubsection{Loss to follow-up with complete
replacement}\label{loss-to-follow-up-with-complete-replacement}}

Here it is convenient to pause and derive the variance of the loss to
follow-up with full replacement estimator. In this case, \(L_i = G_i\)
so equation \ref{eq:vard4a2} becomes \small \begin{align} 
\text{Var}(\hat\beta_3) &= 4\frac{\sigma_{CT}^2}{J} + 2\frac{\sigma_{S}^2}{JK} + 2\frac{\sigma_{ST}^2}{JK} + \frac{\sigma_{S}^2 + \sigma_{ST}^2}{JK} + \frac{\sigma_{S}^2+\sigma_{ST}^2}{JK} - 2\frac{\sigma_{S}^2}{JK}\Big(\frac{K-L_1}{K}+\frac{K-L_2}{K}\Big) \nonumber \\
                        &= 4\frac{\sigma_{CT}^2}{J} + 4\frac{\sigma_{ST}^2}{JK} + 2\frac{\sigma_{S}^2}{JK}\Big(\frac{L_1}{K}+\frac{L_2}{K}\Big) \nonumber \\
                        &= 4\frac{\sigma_{CT}^2}{J} + 4\frac{\sigma_{ST}^2}{JK} + 4\frac{\sigma_{S}^2}{JK}\Big(\frac{\lambda_1+\lambda_2}{2}\Big) \nonumber \\
                        &= 4\Big(\frac{\sigma_{CT}^2}{J} + \frac{\sigma_{ST}^2}{JK} + \frac{\bar\lambda\sigma_{S}^2}{JK}\Big) \nonumber \\
                        &= 4\Big(\frac{\sigma_{CT}^2}{J} + \frac{(1-\rho_S)(\sigma_S^2+\sigma_{ST}^2)+\bar\lambda\rho_S(\sigma_S^2+\sigma_{ST}^2)}{JK}\Big) \nonumber \\
                        &= 4\Big(\frac{\sigma_{CT}^2}{J} + \frac{(1-\rho_S)(\sigma_S^2+\sigma_{ST}^2)+\bar\lambda\rho_S(\sigma_S^2+\sigma_{ST}^2)}{JK}\Big) \nonumber \\
                        &= 4\Big(\frac{\sigma_{CT}^2}{J} + \frac{(1-\rho_S+\bar\lambda\rho_S)(\sigma_S^2+\sigma_{ST}^2)}{JK}\Big) \nonumber \\
                        &= 4\Big(\frac{\sigma_{CT}^2}{J} + \frac{(1-(1-\bar\lambda)\rho_S)(\sigma_S^2+\sigma_{ST}^2)}{JK}\Big) \nonumber \\
                        &= 4\Big(\frac{\sigma_{CT}^2}{J} + \frac{(1-\rho_S^*)(\sigma_S^2+\sigma_{ST}^2)}{JK}\Big)
\end{align} \normalsize

where \(\rho_S^* = (1-\bar\lambda)\rho_S\).

\hypertarget{loss-to-follow-up-with-no-or-partial-replacement}{%
\subsubsection{Loss to follow-up with no or partial
replacement}\label{loss-to-follow-up-with-no-or-partial-replacement}}

In the case of loss to follow-up with no or partial replacement, we
continue from equation \ref{eq:vard4a2} as follows: \small \begin{align}
\text{Var}(\hat\beta_3) &= 4\frac{\sigma_{CT}^2}{J} + 2\frac{\sigma_{ST}^2}{JK} + \frac{\sigma_{S}^2 + \sigma_{ST}^2}{J(K-L_1+G_1)} + \frac{\sigma_{S}^2+\sigma_{ST}^2}{J(K-L_2+G_2)} - 2\frac{\sigma_{S}^2}{JK}\Big(\frac{K-L_1}{K-L_1+G_1}+\frac{K-L_2}{K-L_2+G_2}-1\Big) \nonumber \\
        &= 4\frac{\sigma_{CT}^2}{J} + 4\frac{\sigma_{ST}^2}{JK} -2\frac{\sigma_{ST}^2}{JK} + \frac{\sigma_{S}^2 + \sigma_{ST}^2}{J(K-L_1+G_1)} + \frac{\sigma_{S}^2+\sigma_{ST}^2}{J(K-L_2+G_2)} - 2\frac{\sigma_{S}^2}{JK}\Big(\frac{K-L_1}{K-L_1+G_1}+\frac{K-L_2}{K-L_2+G_2}-1\Big)
\end{align}\normalsize

Letting \(\eta = \frac{K-L_1}{K-L_1+G_1}+\frac{K-L_2}{K-L_2+G_2}-1\), we
get \small \begin{align} \label{eq:vard4a2b}
\text{Var}(\hat\beta_3) &= 4\frac{\sigma_{CT}^2}{J} + 4\frac{\sigma_{ST}^2}{JK} + \frac{(\sigma_{S}^2 + \sigma_{ST}^2)}{J}\Big[\frac{1}{K-L_1+G_1} + \frac{1}{K-L_2+G_2}\Big] - \frac{2(\eta\sigma_S^2 + \sigma_{ST}^2)}{JK} \nonumber \\
        &= 4\frac{\sigma_{CT}^2}{J} + 4\frac{\sigma_{ST}^2}{JK} + \frac{(\sigma_{S}^2 + \sigma_{ST}^2)}{J}\Big[\frac{1}{K-L_1+G_1} + \frac{1}{K-L_2+G_2}\Big] - \frac{2(\eta\sigma_S^2 + \sigma_{ST}^2)}{JK} \nonumber \\
        &= 4\frac{\sigma_{CT}^2}{J} + 4\frac{\sigma_{ST}^2}{JK} + \frac{(\sigma_{S}^2 + \sigma_{ST}^2)}{J}\Big[\frac{1}{K-L_1+G_1} + \frac{1}{K-L_2+G_2}\Big] - \frac{2(\eta\sigma_S^2 + \sigma_{ST}^2)}{JK} \nonumber \\
        &= 4\frac{\sigma_{CT}^2}{J} + 4\frac{\sigma_{ST}^2}{JK} + \frac{(\sigma_{S}^2 + \sigma_{ST}^2)}{JK}\Big[\frac{K}{K-L_1+G_1} + \frac{K}{K-L_2+G_2} - 2\frac{\eta\sigma_S^2 + \sigma_{ST}^2}{\sigma_S^2+\sigma_{ST}^2}\Big] \nonumber \\
        &= 4\frac{\sigma_{CT}^2}{J} + 4\frac{(1-\rho_S)(\sigma_{S}^2+\sigma_{ST}^2)}{JK} + \frac{(\sigma_{S}^2 + \sigma_{ST}^2)}{JK}\Big[\frac{K}{K-L_1+G_1} + \frac{K}{K-L_2+G_2} - 2\frac{\eta\sigma_S^2 + \sigma_{ST}^2}{\sigma_S^2+\sigma_{ST}^2}\Big] \nonumber \\
        &= 4\frac{\sigma_{CT}^2}{J} + 4\frac{(1-\rho_S)(\sigma_{S}^2+\sigma_{ST}^2)}{JK} + 4\frac{1}{4}\frac{(\sigma_{S}^2 + \sigma_{ST}^2)}{JK}\Big[\frac{K}{K-L_1+G_1} + \frac{K}{K-L_2+G_2} - 2\frac{\eta\sigma_S^2 + \sigma_{ST}^2}{\sigma_S^2+\sigma_{ST}^2}\Big] \nonumber \\
        &= 4\frac{\sigma_{CT}^2}{J} + 4\frac{(1-\rho_S)(\sigma_{S}^2+\sigma_{ST}^2)}{JK} + 4\frac{1}{4}\frac{(\sigma_{S}^2 + \sigma_{ST}^2)}{JK}\Big[\frac{K}{K-L_1+G_1} + \frac{K}{K-L_2+G_2} - 2\frac{\eta\sigma_S^2 + \sigma_{ST}^2}{\sigma_S^2+\sigma_{ST}^2}\Big] \nonumber \\
        &= 4\frac{\sigma_{CT}^2}{J} + 4\Bigg(1-\rho_S + \frac{1}{4}\Big[\frac{1}{1-\lambda_1+\gamma_1} + \frac{1}{1-\lambda_2+\gamma_2} - 2\frac{\eta\sigma_S^2 + \sigma_{ST}^2}{\sigma_S^2+\sigma_{ST}^2}\Big]\Bigg)\frac{(\sigma_{S}^2 + \sigma_{ST}^2)}{JK} 
\end{align} \normalsize

where \(\lambda_i = L_i/K\) is the loss to follow-up per cluster in
group \(i\) and \(\gamma_i = G_i/K\) is the gain at follow-up (relative
to the initial cluster size, \(K\)) per cluster in group \(i\).

If
\(\rho_S* = \rho_S - \frac{1}{4}\Big[\frac{1}{1-\lambda_1+\gamma_1} + \frac{1}{1-\lambda_2+\gamma_2} - 2\frac{\eta\sigma_S^2 + \sigma_{ST}^2}{\sigma_S^2+\sigma_{ST}^2}\Big]\),
then we can write

\begin{align}
\text{Var}(\hat\beta_3) &= 4\frac{\sigma_{CT}^2}{J} + 4\frac{(1-\rho_S^*)(\sigma_S^2+\sigma_{ST}^2)}{JK}
\end{align}

\newpage

\hypertarget{references}{%
\section*{References}\label{references}}
\addcontentsline{toc}{section}{References}

\hypertarget{refs}{}
\leavevmode\hypertarget{ref-feldman1994}{}%
Feldman, Henry A., and Sonja M. McKinlay. 1994. ``Cohort Versus
Cross-Sectional Design in Large Field Trials: Precision, Sample Size,
and a Unifying Model.'' \emph{Statistics in Medicine} 13 (1):61--78.
\url{https://doi.org/10.1002/sim.4780130108}.

\leavevmode\hypertarget{ref-fiero2016}{}%
Fiero, Mallorie H., Shuang Huang, Eyal Oren, and Melanie L. Bell. 2016.
``Statistical Analysis and Handling of Missing Data in Cluster
Randomized Trials: A Systematic Review.'' \emph{Trials} 17 (72).
\url{https://doi.org/10.1186/s13063-016-1201-z}.

\leavevmode\hypertarget{ref-hayesmoulton2017}{}%
Hayes, Richard J., and Lawrence H. Moulton. 2017. \emph{Cluster
Randomized Trials}. 2nd ed. CRC Press.

\leavevmode\hypertarget{ref-reducemrsa}{}%
Huang, Susan S., Edward Septimus, Ken Kleinman, Julia Moody, Jason
Hickok, Taliser R. Avery, Julie Lankiewicz, et al. 2013. ``Targeted
Versus Universal Decolonization to Prevent Icu Infection.'' \emph{New
England Journal of Medicine} 368 (24):2255--65.
\url{https://doi.org/10.1056/NEJMoa1207290}.

\leavevmode\hypertarget{ref-clusterPower}{}%
Kleinman, Ken, Jon Moyer, and Nicholas Reich. 2017. \emph{ClusterPower:
Power Calculations for Cluster-Randomized and Cluster-Randomized
Crossover Trials}.
\url{https://CRAN.R-project.org/package=clusterPower}.

\leavevmode\hypertarget{ref-miller2012}{}%
Miller, Elizabeth, Daniel J Tancredi, Heather L McCauley, Michele R
Decker, Maria Catrina D Virata, Heather A Anderson, Nicholas Stetkevich,
Ernest W Browne, Feroz Moideen, and Jay G Silverman. 2012. ``Coaching
Boys into Men: A Cluster-Randomized Controlled Trial of a Dating
Violence Prevention Program.'' \emph{Journal of Adolescent Health} 51
(5):431--38. \url{https://doi.org/10.1016/j.jadohealth.2012.01.018}.

\leavevmode\hypertarget{ref-murray1990}{}%
Murray, David M., and Peter J. Hannan. 1990. ``Planning for the
Appropriate Analysis in School-Based Drug-Use Prevention Studies.''
\emph{Journal of Consulting and Clinical Psychology} 58 (4):458--68.
\url{https://doi.org/http://dx.doi.org/10.1037/0022-006X.58.4.458}.

\leavevmode\hypertarget{ref-pettifor2018}{}%
Pettifor, Audrey, Sheri A. Lippman, Ann Gottert, Chirayath M.
Suchindran, Selin Amanda, Dean Peacock, Suzanne Maman, et al. 2018.
``Community Mobilization to Modify Harmful Gender Norms and Reduce Hiv
Risk: Results from a Community Cluster Randomized Trial in South
Africa.'' \emph{Journal of the International AIDS Society} 21
(7):e25134. \url{https://doi.org/10.1002/jia2.25134}.

\leavevmode\hypertarget{ref-rubin1976}{}%
Rubin, Donald B. 1976. ``Inference and Missing Data.'' \emph{Biometrika}
63 (1):581--92. \url{https://doi.org/10.1093/biomet/63.3.581}.

\leavevmode\hypertarget{ref-rutterford2015}{}%
Rutterford, Clare, Andrew Copas, and Sandra Eldridge. 2015. ``Methods
for Sample Size Determination in Cluster Randomized Trials.''
\emph{International Journal of Epidemiology} 44 (3):1051--67.
\url{https://doi.org/10.1093/ije/dyv113}.

\leavevmode\hypertarget{ref-strasser2008}{}%
Strasser, Dale C, Judith A Falconer, Alan B Stevens, Jay M Uomoto, Jeph
Herrin, Susan E Bowen, and Andrea B Burridge. 2008. ``Team Training and
Stroke Rehabilitation Outcomes: A Cluster Randomized Trial.''
\emph{Archives of Physical Medicine and Rehabilitation} 89 (1):10--15.
\url{https://doi.org/10.1016/j.apmr.2007.08.127}.

\leavevmode\hypertarget{ref-teerenstra2012}{}%
Teerenstra, Steven, Sandra Eldridge, Maud Graff, Ester deHoop, and
George F. Born. 2012. ``A Simple Sample Size Formula for Analysis of
Covariance in Cluster Randomized Trials.'' \emph{Statistics in Medicine}
31 (20):2169--78.
\href{https://doi.org/10.1002/sim.5352\%20}{https://doi.org/10.1002/sim.5352}.

\end{document}